\documentclass[submitting]{nst}

\usepackage{subfigure,dcolumn}
\usepackage{epstopdf}
\usepackage{mhchem}
\usepackage{upgreek}
\usepackage{color}

\begin{document}

\title{A simulation study of a windowless gas stripping room in an E//B neutral particle analyzer}
\thanks{Supported by the National MCF Energy R\&D Program of China (MOST 2018YFE0310200), the National Natural Science Foundation of China (Grant No. 11805138 and 11705242) and the Fundamental Research Funds For the Central Universities (No. YJ201820 and YJ201954).}

\author{Yuan Luo}
\affiliation{Key Laboratory of Radiation Physics and Technology of the Ministry of Education, Institute of Nuclear Science and Technology, Sichuan University, Chengdu 610064, China}
\author{Wei-Ping Lin}
\email[Corresponding author, ]{linwp1204@scu.edu.cn.}
\affiliation{Key Laboratory of Radiation Physics and Technology of the Ministry of Education, Institute of Nuclear Science and Technology, Sichuan University, Chengdu 610064, China}
\author{Pei-Pei Ren}
\affiliation{Key Laboratory of Radiation Physics and Technology of the Ministry of Education, Institute of Nuclear Science and Technology, Sichuan University, Chengdu 610064, China}
\author{Guo-Feng Qu}
\affiliation{Key Laboratory of Radiation Physics and Technology of the Ministry of Education, Institute of Nuclear Science and Technology, Sichuan University, Chengdu 610064, China}
\author{Jing-Jun Zhu}
\email[Corresponding author, ]{zhujingjun@scu.edu.cn.}
\affiliation{Key Laboratory of Radiation Physics and Technology of the Ministry of Education, Institute of Nuclear Science and Technology, Sichuan University, Chengdu 610064, China}
\author{Xing-Quan Liu}
\affiliation{Key Laboratory of Radiation Physics and Technology of the Ministry of Education, Institute of Nuclear Science and Technology, Sichuan University, Chengdu 610064, China}
\author{Xiao-Bing Luo}
\affiliation{Key Laboratory of Radiation Physics and Technology of the Ministry of Education, Institute of Nuclear Science and Technology, Sichuan University, Chengdu 610064, China}
\author{Zhu An}
\affiliation{Key Laboratory of Radiation Physics and Technology of the Ministry of Education, Institute of Nuclear Science and Technology, Sichuan University, Chengdu 610064, China}
\author{Roy Wada}
\affiliation{Cyclotron Institute, Texas A$\&$M University, College Station, Texas 77843, USA}
\affiliation{School of Physics, Henan Normal University, Xinxiang 453007, China}
\author{Lin-Ge Zang}
\affiliation{Southwestern Institute of Physics, Chengdu 610064, China}
\author{Yu-Fan Qu}
\affiliation{Southwestern Institute of Physics, Chengdu 610064, China}
\author{Zhong-Bing Shi}
\affiliation{Southwestern Institute of Physics, Chengdu 610064, China}

\begin{abstract}
Neutral Particle Analyzer (NPA) is one of the crucial diagnostic devices on Tokamak facilities. Stripping unit is one of the main parts of the NPA. A windowless gas stripping room with two differential pipes is adopted in a parallel direction of electric and magnetic fields (E//B) NPA. The pressure distributions in the stripping chamber are simulated by Ansys Fluent together with MolFlow+. Based on the pressure distributions extracted from the simulation, the stripping efficiency of the E//B NPA is studied with GEANT4. The hadron reaction physics is modified to track the charge state of each particle in a cross section base method in GEANT4. The transmission rates ($R$) and the stripping efficiencies $f_{+1}$ are examined for the particle energy ranging from 20 to 200 keV at the input pressure ($P_0$) ranging from 20 to 400 Pa. According to the combined global efficiency, $R \times f_{+1}$, $P_0$ = 240 Pa is obtained as the optimum pressure for the maximum global efficiency in the incident energy range investigated.
\end{abstract}


\maketitle

\section{Introduction}\label{sec.I}

Tokamak is a toroidal device used in nuclear fusion research for the magnetic confinement plasma. It provides a place to test the integrated technologies, materials, and physics regimes necessary for the future commercial production of fusion-based electricity~\cite{ITERwebsite}. 
Neutral Particle Analyzer (NPA) is one of the crucial diagnostic devices on Tokamak facilities. It is used to determine the bulk ion temperature, the isotopic ratio and the fast ion distribution of the plasma, by measuring the charge exchange neutral particles escaping from the plasma. 
Different types of NPA have been built in Tokamak facilities worldwide~\cite{Medley1998RSI,Chernyshev2004IET,Petrov2017PAN,Isobe2001RSI,Krasilnikov2008IET,Bracco1992RSI,Liu2016RSI,Afanasyev2003RSI,Afanasyev2010NIMA,Zhang2016RSI,Li2012RSI,Bartiromo1987RSI}, such as the parallel direction of electric and magnetic fields (E//B) NPA on the Tokamak Fusion Test Reactor (TFTR)~\cite{Medley1998RSI}, the compact neutral particle analyzer (CNPA) on the Wendelstein 7-AS
stellarator~\cite{Chernyshev2004IET}, the low- and high- energy neutral particle analyzers (LENPA and HENPA) on the International Thermonuclear Experimental Reactor (ITER)~\cite{Afanasyev2010NIMA}, the solid state NPA (ssNPA) on the Experimental Advanced Superconducting Tokamak (EAST)~\cite{Zhang2016RSI} and the CP-NPA on the HuanLiuqi-2A (HL-2A)~\cite{Li2012RSI}.

Stripping unit plays an important role in the analyzing of the neutral particles, except for the flux measurement NPA such as ssNPA~\cite{Zhang2016RSI}. It provides a place to reionize the charge exchange neutral particles. According to the state of the stripping material, the stripping unit can be separated into two type, the stripping foil and the gas chamber. When a stripping foil is used in the NPA for the low energy neutrals, an additional accelerating or focusing voltage is required for the secondary ions~\cite{Afanasyev2003RSI,Chernyshev2004IET,Afanasyev2010NIMA}. A carbon foil with the thickness of 100 \r{A} is commonly used as the stripping foil. On the contrary, a gas chamber requires a differential pumping system when the stripping gas is used. Typically the integrated target thickness of the order of $10^{16}$ atoms/cm$^{2}$ for the H$_2$ gas is used in the Joint European Torus (JET) NPA~\cite{Bartiromo1987RSI}, and $10^{15}$ atoms/cm$^{2}$ for the He gas is used in the E//B NPA on TFTR~\cite{Medley1998RSI}.

The energetic particles, also known as fast, superthermal,	hot and high-energy particles, are expected to play a critical role in plasma heating, current drive, momentum transport, energy transfer and plasma stability~\cite{WChen2020CPL,LChen2016RMP}. Many experimental and theoretical studies have been contributed to this field~\cite{Shi2021CPL,YChen2020CPL,Madsen2020PPCF,Gorelenkov2014NF,Ding2018PST}, and other related fields~\cite{Wei2021CPL,Xu2020CPL,Wan2020CPL,Zheng2019NST1,Shu2019NST,Cui2019NST} recently. Aiming to study the frontier physics of the energetic particles and to measure the fuel ratio, a new E//B NPA has been designed in the present experimental devices~\cite{Zang2020ITC-28}. This E//B NPA is a tandem type NPA like CNPA built in Ioffe Physicotechnical Institute, Russia~\cite{Chernyshev2004IET}. It will provide mass resolution (H and D resolution) for particles in the energy range of 20 to 200 keV. The magnetic field is designed to be created with a permanent magnet for smaller size and simpler maintenance. The upper limit energy of the E//B NPA is determined from the negative ion source neutral beam heating on Huanliuqi-2M (HL-2M) device. The lower limit is set to 20 keV because we are interested in the fast ions, instead of the background ions. For more details, we refer to our previous work in Ref.~\cite{Zang2020ITC-28}.

In this article, the gas stripping chamber of the new E//B NPA is designed and studied. A windowless gas stripping chamber is adopted to avoid the replacement of the stripping foils and make an easy maintenance in the actual operation. The performance of the gas stripping chamber is investigated using Ansys Fluent~\cite{AnsysFluent,Mohamedi2015NST} and MolFlow+~\cite{MolFlow}, together with GEANT4~\cite{Agostinelli2003GEANT4,Allison2016GEANT4}. This article is organized as follows: The design and pressure calculation of the gas stripping chamber is presented in Sec.~\ref{sec.II}. The results of GEANT4 simulation and discussions are given in Sec.~\ref{sec.III}. A brief summary is given in Sec.~\ref{sec.IV}.

\section{Design and pressure calculation of the gas stripping chamber}\label{sec.II}

\begin{figure}[!hbt]
	\centering
	\includegraphics[width=\hsize]{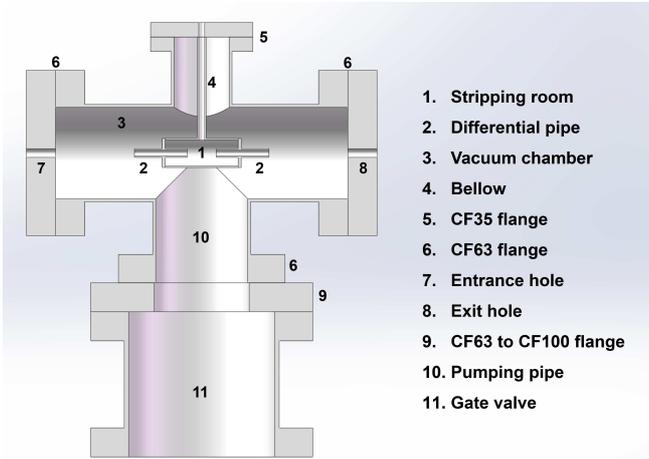}
	\caption{\footnotesize
		Schematic diagram of the gas stripping chamber.
	}
	\label{fig:fig01_layout}
\end{figure}

Stripping unit is one of the main parts of the NPA. The electron of the escaped neutral particles will be stripped in the stripping unit. A windowless gas stripping chamber is adopted in the design to avoid the replacement of the stripping foil and for simple maintenance. In order to get a certain high pressure inside the stripping room but high vacuum in the outside vacuum chamber at the same time, two differential pipes with small flow conductance are used for stripping room. Fig.~\ref{fig:fig01_layout} shows the schematic layout of the stripping chamber. The stripping room (1) with two differential pipes (2) of 36 mm in length and 4 mm in diameter is placed inside a vacuum chamber (3). Two holes with a diameter of 6 mm (7 and 8) are made on the entrance and exit flanges to limit the beam size and to maximize the vacuum isolation from the upstream pipe and downstream chamber. H$_2$ gas is used as the stripping gas to avoid polluting the Tokamak fuel. It is filled in the stripping room from the top flange (5) of the vacuum chamber through a bellow (4). A machinery bearing molecular pump with a pumping speed of 340 L/s is used at the bottom of vacuum chamber together with a gate valve (11) in this work. 

The pressure distribution inside the stripping chamber is one of the main concerns of our design. The pressure of dozens of Pa is required in the stripping room to get enough stripping efficiency for high energy hydrogen (H) and deuteron (D) atoms. At this pressure region, the gas flow state in the stripping room stays at viscous-molecular flow~\cite{Da2004Zhenkong}. The pressure in the gas inlet and the bellow is higher than that in the stripping room. On the contrary, two or three orders lower pressure is estimated in the vacuum chamber. The mean free path of the gas molecular inside the vacuum chamber is larger than the size of chamber and the motion of the gas molecular can be treated collisionlessly. The Monte Carlo code of MolFlow+ is often used to calculate the pressure distribution of the collisionless gas in high vacuum system. But it is not accurate for all the gas region in the gas stripping chamber. The computational fluid dynamics (CFD) software, which includes the nonlinear effect of viscous fluid, has better performance in the high pressure region. However, the calculation of the CFD software at high vacuum region shows unphysical bump at corners. Therefore, the gas pressure distribution of the stripping chamber is calculated, combining a CFD software of Ansys Fluent~\cite{AnsysFluent,Mohamedi2015NST} for the viscous region in the bellow, stripping room, and differential pipes and a Monte Carlo software of MolFlow+~\cite{MolFlow} for the low pressure collisionless region in the vacuum chamber.

Three dimensional CFD calculations are performed in the Ansys Fluent software. The fluid region is established according to the structure of the stripping chamber shown in Fig.~\ref{fig:fig01_layout}. The laminar vicious model is adopted in the calculations. A pressure type gas inlet is defined at the top flange (5) of vacuum chamber and the pressure at the gas inlet $P_{0}$ ranges from 20 to 400 Pa with step of 20 Pa set in the simulation. Three pressure type gas outlets are defined at the entrance, exit holes (7 and 8) and the bottom of gate valve (11). A pressure of $10^{-3}$ Pa is assumed for all three outlets. Due to the large flow conductance of the two differential pipes (2), small change in the outlet pressure does not affect the pressure distribution in the stripping room. Moreover, the pressure distribution in the vacuum chamber will be replaced with the results of MolFLow+. Therefore, the pressure of $10^{-3}$ Pa of the outlets is used for all the Ansys Fluent calculations. The stainless steel is set as the wall material. The room temperature of 300 K is used in the calculations. A typical gas flow rate of 9.97 Pa$\cdot$L/s is obtained at the bottom of the bellow for the input pressure $P_0$ = 100 Pa.

In the low pressure region in the vacuum chamber (3) in Fig.~\ref{fig:fig01_layout} and inside the differential pipe (2) at $|z|>$ 42 mm, where z = 0 is set to the central of the stripping room, the pressure distribution is simulated by MolFlow+ at the same temperature of 300 K. For the MolFlow+ simulation, the outgassing rate adopted is from the Ansys Fluent calculation at z $=\pm$42 mm inside the differential pipe, 4 mm from the pipe exit. For the pumping, it is assumed that the gas molecules are absorbed when they hit the surfaces of the entrance and exit holes (7 and 8), that is, the sticking factor is set to 1 on the surfaces. This results in the pumping speed of 12.4 L/s through the entrance and exit holes. A pumping speed of 340 L/s is set at the bottom of the gate valve (11).

\begin{figure}[!hbt]
	\centering
	\includegraphics[width=\hsize]{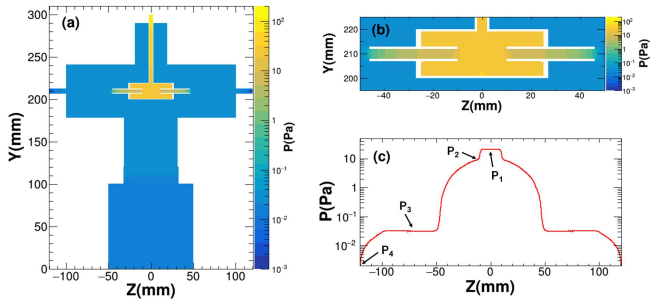}
	\caption{\footnotesize
		(a) 2D pressure distribution in the y-z plane at x = 0 for the combined simulation of Ansys Fluent and MolFlow+ at $P_0=$ 100 Pa. (b) Magnified pressure distribution around the stripping room. (c) Pressure distribution along the beam line as a function of z in a logarithmic scale. $P_1$, $P_2$, $P_3$ and $P_4$ indicate four typical pressures in the stripping chamber.
	}
	\label{fig:fig02_AnsysFluent_MolFlow}
\end{figure}

The simulated two dimensional (2D) pressure distribution at $P_0=$ 100 Pa in the y-z plane at x = 0 is shown in Fig.~\ref{fig:fig02_AnsysFluent_MolFlow} (a). A detailed pressure distribution around the stripping room is shown in Fig.~\ref{fig:fig02_AnsysFluent_MolFlow} (b) in a magnified scale. The pressure distribution is presented along the beam line in Fig.~\ref{fig:fig02_AnsysFluent_MolFlow} (c) plotted in a logarithmic scale. Using the two differential pipes in the design, the desired high pressure is achieved inside the stripping room, and the linear decreasing pressure is observed inside the two differential pipes and the entrance and exit holes. A sharp change of the pressure at the entrance of the two differential pipes is observed in the Ansys Fluent calculations. It is not found when MolFlow+ is used to simulate the whole gas region in the stripping chamber.

\begin{figure}[!hbt]
	\centering
	\includegraphics[scale=0.4]{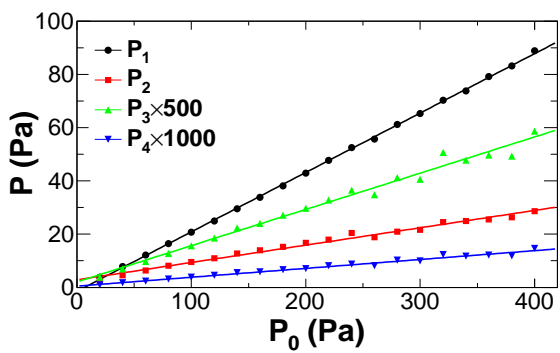}
	\caption{\footnotesize
		Pressure $P_1$ (solid circles), $P_2$ (solid squares), $P_3$ (solid up triangles) and $P_4$ (solid down triangles) as a function of the $P_0$. The lines are obtained by the linear fits.
	}
	\label{fig:fig03_PvsP0}
\end{figure}

To evaluate the pressure distribution changed as the pressure at the gas inlet changes, the pressures at four typical positions are examined for all the $P_0$ investigated. Fig.~\ref{fig:fig03_PvsP0} shows the pressure inside the stripping room ($P_1$), at entrance of the differential pipes ($P_2$), at the vacuum chamber ($P_3$) and at the outside surfaces of entrance and exit holes ($P_4$), which is pointed out in Fig.~\ref{fig:fig02_AnsysFluent_MolFlow} (c), as a function of the pressure at the gas inlet $P_0$. Linear changes of $P_1$, $P_2$, $P_3$ and $P_4$ on $P_0$ are obtained, though slight fluctuations are found. By using the two differential pipes in the design, more than 500 times lower pressure is observed in the vacuum chamber than that in the stripping room. Through the entrance and exit holes, about one order magnitude lower pressure is obtained for the upstream pipe and downstream chamber.

\begin{figure}[!hbt]
	\centering
	\includegraphics[scale=0.4]{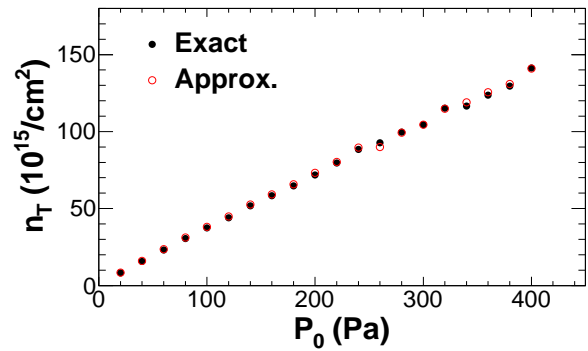}
	\caption{\footnotesize
		$n_T$ as a function of $P_0$. Solid circles are the exact $n_T$ calculated from the pressure distributions of the Ansys Fluent and the MolFlow+ simulations. Open circles are the $n_T$ calculated from the pressure distributions constructed in the GEANT4 simulation.
	}
	\label{fig:fig04_target_thickness}
\end{figure}

The integrated target thickness ($n_{T}$) is an important quantity of the gas stripping chamber. It is commonly used to evaluate the efficiency of a stripping chamber. Since the pressures at the four typical positions are used to construct the pressure distribution in GEANT4 in the next section, a comparison of the exact $n_{T}$ and the $n_{T}$ calculated from the pressure distribution used in GEANT4 is necessary. Fig.~\ref{fig:fig04_target_thickness} shows $n_T$ as a function of $P_0$ for the results of Ansys Fluent and MolFlow+ (solid circles) and that of GEANT4 (open circles). Good agreement is found between them.

\section{Results of GEANT4 simulation and discussions}\label{sec.III}

Utilizing the obtained $P_1$, $P_2$, $P_3$ and $P_4$, the stripping efficiencies of the stripping room are further studied using GEANT4~\cite{Agostinelli2003GEANT4,Allison2016GEANT4}. GEANT4 is a toolkit for the simulation of the passage of particles through matter. It has been applied for varies studies, such as our previous studies of the average neutron detection efficiency for the DEtecteur MOdulaire de Neutrons detectors (DEMON)~\cite{Zhang2013NIMA} and the module test of the Collision Centrality Detector Array (CCDA)~\cite{Fen2014CPL}, and for many other studies of the electron backscattering~\cite{Dondero2018NIMB}, the neutron time of flight spectrometer system at HL-2M~\cite{Zheng2019NST}, the performance of a large size CsI detector~\cite{Dong2018NST} and so on.

In the GEANT4 simulation, the physics list includes the electromagnetic physics~\cite{Incerti2018EMphysics} and the hadronic physics~\cite{Wellisch2001Hadronic}, in which ion transportation, electromagnetic, nuclear elastic and inelastic processes are activated, though some of the processes may not be used in the actual simulation. Since the integrated target thickness is around $10^{16}$ atoms/cm$^{2}$, in which the scattering probability of the incident of particles and the target atoms is small enough that the multi-scattering is negligible. Therefore, the G4ScreenedNuclearRecoil class~\cite{Weller2004G4Screened,Mendenhall2005G4CoulombScattering} is included in the standard electromagnetic physics for the incident energy ranging from 10 eV to 100 MeV. 

\begin{figure}[hbt]
	\centering
	\includegraphics[width=\hsize]{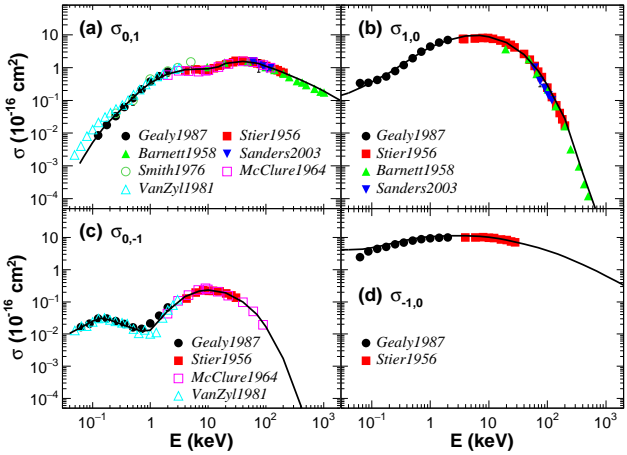}
	\caption{\footnotesize
		Charge exchange cross section of $\sigma_{0,1}$ (a),  $\sigma_{1,0}$ (b),  $\sigma_{0,-1}$ (c) and  $\sigma_{-1,0}$ (d) in unit of $10^{-16} $cm$^{2}$ per molecule as a function of the incident energy ($E$) of H atom or ion. Solid circles, solid squares, solid up triangles, solid down triangles, open circles, open squares and open up triangles represent the data of Gealy~\cite{Gealy1987PRA}, Stier~\cite{Stier1956PR}, Barnett~\cite{Barnett1958PR}, Sanders~\cite{Sanders2003JPB}, Smith~\cite{Smith1976JGR}, McClure~\cite{McClure1964PR} and Van Zyl~\cite{VanZyl1981}, respectively. Solid curves represent the ORNL recommended cross section~\cite{Hunter1990ADF,Schultz2020ADNDT}, which is also used in the simulation.
	}
	\label{fig:fig05_charge_exchange_cross_section}
\end{figure}

The charge state of H and D atoms is the key variable in this study. However, the original GEANT4 cannot handle the charge state properly. In order to simulate the charge state variation, the hadron reaction physics is modified to trace the charge state of each particle in a cross section base method. By introducing a global charge state variable in the hadron reaction physics, the charge state of H and D is recorded when the charge exchange reaction happened in the gas stripping chamber. Many charge exchange cross section measurements have been performed for H on H$_2$ gas during the last century~\cite{Barnett1958PR,Curran1960PR,Gealy1987PRA,McClure1964PR,Sanders2003JPB,Sigaud2011JPB,Smith1976JGR,Stier1956PR,VanZyl1981,Hunter1990ADF,Schultz2020ADNDT}. Fig.~\ref{fig:fig05_charge_exchange_cross_section} shows the electron loss cross sections of $H^{0}$ ($\sigma_{0,1}$), the electron capture cross sections of  $H^{+}$ ($\sigma_{1,0}$), the electron capture cross sections of  $H^{0}$ ($\sigma_{0,-1}$) and the electron loss cross sections of $H^{-}$ ($\sigma_{-1,0}$) on H$_2$ gas as a function of the incident energy ($E$) in (a), (b), (c) and (d), respectively. Solid circles, solid squares, solid up triangles, solid down triangles, open circles, open squares and open up triangles represent the data of Gealy~\cite{Gealy1987PRA}, Stier~\cite{Stier1956PR}, Barnett~\cite{Barnett1958PR}, Sanders~\cite{Sanders2003JPB}, Smith~\cite{Smith1976JGR}, McClure~\cite{McClure1964PR} and Van Zyl~\cite{VanZyl1981}, respectively. Solid curves represent the ORNL recommended cross section~\cite{Hunter1990ADF,Schultz2020ADNDT}. Since the experimental $\sigma_{-1,0}$ does not cover the high energy region at $E$ greater than 30 keV, the ORNL recommended cross section of H atom on H$_2$ gas is used in the simulation. For a given velocity, the charge exchange cross sections of D on Cs~\cite{Meyer1975PLA} or Rb~\cite{Schlachter1969PR} vapor are the same as those of H. Approximately, the incident energy per nucleon ($E/A$) of H and D are the same for the energy range investigated when they have the same velocity. Therefore, the charge exchange cross sections of H are also used for D at the same $E/A$ in the simulations.

\begin{figure}[!htb]
	\centering
	\includegraphics[scale=0.35]{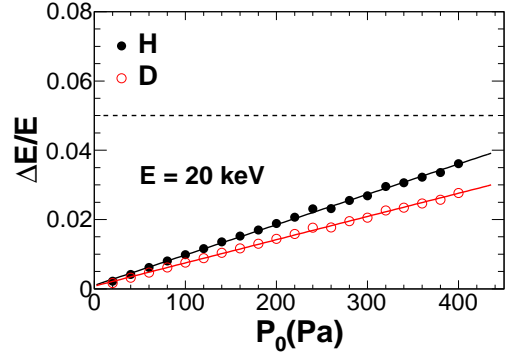}
	\caption{\footnotesize
		Energy loss of 20 keV H and D passing though the stripping chamber as a function of $P_0$. Dashed line shows as the guide line of $\Delta E/E$ = 0.05. Solid lines are from the linear fits.
	}
	\label{fig:fig06_dE_vs_P0}
\end{figure}

The simulations are performed for H and D on H$_{2}$ gas with the incident energy ranging from 20 to 200 keV in the step of 20 keV, and with $P_{0}$ ranging from 20 to 400 Pa in the step of 20 Pa. H and D atoms are generated at the entrance hole (7) of the vacuum chamber, corresponding to the z position of -120 mm, and distributed uniformly on the entrance hole surface with the same diameter of 6 mm. The momentum direction is assumed parallel to the z-axis. One million events are generated for each run. The energy loss ($\Delta E/E$) of H and D at 20 keV as a function of $P_0$ is shown in Fig.~\ref{fig:fig06_dE_vs_P0}. A slight lower energy loss is observed for D, because the mass of D is twice of that of H, which causes less energy loss during the collisions. One can see from Fig.~\ref{fig:fig06_dE_vs_P0} that the linear increasing trends are found for H and D as $P_0$ increases. The maximum energy loss of 20 keV H and D is less than 4\% for all $P_0$ investigated. Comparing to the energy resolution of this NPA, the energy loss of H and D after passing through the stripping chamber is small and can be neglected.

\begin{figure}[!htb]
	\centering
	\includegraphics[width=\hsize]{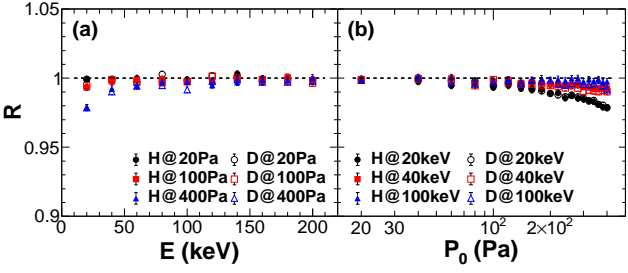}
	\caption{\footnotesize
		(a) $R$ as a function of $E$. Circles, squares and triangles represent $P_{0}$ = 20, 100 and 400 Pa, respectively. (b) $R$ as a function of $P_{0}$. Circles, squares and triangles represent those at $P_{0}$ = 20, 40 and 100 keV, respectively. Solid and open symbols represent for H and D, respectively. Dashed lines show for $R$ = 1. The errors shown for the data points are the statistical errors.
	}
	\label{fig:fig07_Ratio_for_scattering}
\end{figure}

Since the diameter of the two differential pipes is smaller than that of the entrance and the exit holes, some of the incident particles will be stopped by the geometry of the stripping chamber. Moreover, due to the Coulomb scattering of the incident particles and the target atoms, some of the incident particles will be scattered away from their original directions. Therefore, the transmission rate ($R$) of the incident particles is important, especially for low energy particles which suffer more Coulomb scattering, when they pass though the stripping chamber. In this study, the transmission rate is defined as the ratio between the number of particles reached at the exit hole (8) after passing through the stripping chamber with the H$_2$ gas ($P_0 >$ 0 Pa) and that without the gas ($P_0=$ 0 Pa, vacuum). In this way, the particle loss caused by the geometry of the stripping chamber is canceled out. Fig.~\ref{fig:fig07_Ratio_for_scattering} shows the transmission rate as a function of the incident energy for $P_0$ = 20 Pa (circles), 100 Pa (squares) and 400 Pa (triangles) in (a) and as a function of $P_{0}$ for the incident energy E = 20 keV (circles), 40 keV (squares) and 100 keV (triangles) in (b). Solid and open symbols represent H and D, respectively. A slight increasing trend is observed for the transmission rate as the incident energy increases, but show an opposite trend as $P_0$ increases. The scattering loss is small (less than 3\%) for all the incident energies and the input pressures investigated.

\begin{figure}[hbt]
	\centering
	\includegraphics[scale=0.4]{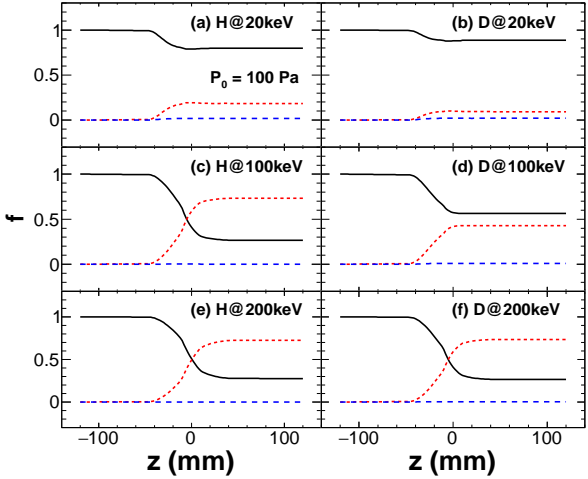}
	\caption{\footnotesize
		Charge state fraction as a function of z position in the stripping chamber for H at 20 keV in (a), D at 20 keV in (b), H at 100 keV in (c), D at 100 keV in (d), H at 200 keV in (e) and D at 200 keV in (f). The pressures used in the simulation are those from the results of $P_{0}$ = 100 Pa. Solid, dashed and long-dashed curves corresponding to fraction of charge state 0, +1 and -1, respectively.
	}
	\label{fig:fig08_Qfraction}
\end{figure}

The stripping efficiency in the stripping chamber is evaluated using a charge fraction variable (f) for the incident particles after the stripping area. The evolution of the charge fraction inside the stripping area is the most concerned in this study. Fig.~\ref{fig:fig08_Qfraction} shows the charge state fraction as a function of z position in the stripping chamber for H and D atoms at 20, 100 and 200 keV. The pressures used in the simulation are those from the results of $P_{0}$ = 100 Pa as shown in Sec.~\ref{sec.II}. Solid, dashed and long-dashed curves correspond to the fraction of the charge state 0, +1 and -1, respectively. Due to the small cross section of $\sigma_{0,-1}$, the fraction of charge state -1 is less than 2\% for H and D at the incident energy of 20 keV and becomes negligible for higher incident energies. The fraction of charge state 0 and +1 show a sharp change started from z around -40 mm, which is because the stripping room locates at -46 mm $<z<$ 46 mm. Due to the energy dependence of the charge exchanging cross section, particles with lower incident energies have smaller saturation thickness for the charge fractions. A higher stripping efficiency for particles with larger energies is mainly caused by the sharp decrease of $\sigma_{1,0}$ as the incident energy increases.

\begin{figure}[!hbt]
	\centering
	\includegraphics[width=\hsize]{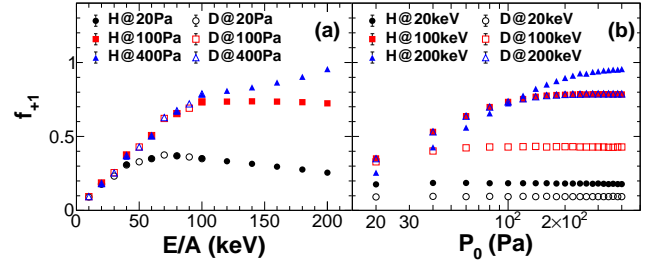}
	\caption{\footnotesize
		(a) $f_{+1}$ as a function of $E/A$. Circles, squares and triangles represent the values at $P_{0}$ = 20, 100 and 400 Pa, respectively. (b) $f_{+1}$ as a function of $P_0$. Circles, squares and triangles represent those at $E$ = 20, 100 and 200 keV, respectively. Solid and open symbols represent H and D, respectively.
	}
	\label{fig:fig09_Q1fraction}
\end{figure}

Figure~\ref{fig:fig09_Q1fraction} shows the fraction of charge state +1 ($f_{+1}$) at z = 120 mm as a function of $E/A$ for $P_0$ = 20 Pa (circles), 100 Pa (squares) and 400 Pa (triangles) in (a), and as a function of $P_{0}$ for the incident energy $E$ = 20 keV (circles), 100 keV (squares) and 200 keV (triangles) in (b). Solid and open symbols represent H and D, respectively. One can see from Fig.~\ref{fig:fig09_Q1fraction} (a) that $f_{+1}$ increases at lower incident energies for different $P_{0}$ up to $E/A$ around 100 keV. After reaching the maximum value, $f_{+1}$ decreases for $P_{0}$ = 20 Pa, and stay flat for $P_0$ = 100 Pa, but keep slowly increasing for $P_{0}$ = 400 Pa as $E/A$ increases, which indicates that the thickness of the stripping gas is not enough for higher energy particles at $P_{0}$ = 20 Pa. No noticeable difference between H and D is observed, indicating that the results can be applied also to neutral Tritium particles when $E/A$ is used. As shown in Fig.~\ref{fig:fig09_Q1fraction} (b), the fractions of charge state +1 quickly reach a maximum value and keep the maximum as $P_{0}$ increases at lower $E$. For larger $E$, the fractions increase faster at lower $P_{0}$ and reach the maximum at pressure around $P_{0}$ = 240 Pa.

\begin{figure}[!hbt]
	\centering
	\includegraphics[scale=0.35]{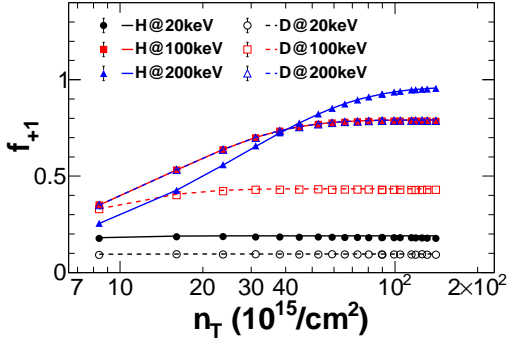}
	\caption{\footnotesize
		The fraction of $H^{+}$ (solid symbols) and $D^{+}$ (open symbols) as a function of $n_T$ for $E$ = 20 keV (circles), 100 keV (squares) and 200 keV (triangles). Solid and dashed lines are the results of Eq.~(\ref{eq:simple_formula}). See more details in the text.
	}
	\label{fig:fig10_fraction_area_density}
\end{figure}

In order to verify the GEANT4 results, $f_{+1}$ is also calculated, using the gas integrated target thickness ($n_T$) as
\begin{equation}\label{eq:simple_formula}
	f_{+1} = \frac{\sigma_{01}}{\sigma_{01}+\sigma_{10}}\left\lbrace 1-\exp{\left[ -n_{T}(\sigma_{01}+\sigma_{10})\right]}\right\rbrace. 
\end{equation} 
The $\sigma_{01}$ and $\sigma_{10}$ are the stripping (the electron loss) and the charge exchange (the electron capture) cross sections of H(D) and H$^{+}$(D$^{+}$), respectively. The small amount of particle loss by the electron capture of H(D), $\sigma_{0-1}$ and $\sigma_{-10}$, is neglected in Eq.~(\ref{eq:simple_formula}). The results are shown in Fig.~\ref{fig:fig10_fraction_area_density}. The symbols are the same $f_{+1}$ values from the GEANT4 simulation in Fig.~\ref{fig:fig09_Q1fraction} (b) but plotted as a function of $n_T$, and solid and dashed lines are those from Eq.~(\ref{eq:simple_formula}) for H and D, respectively. Good agreements are found between the calculations and the GEANT4 simulations. These good agreements originate from the following fact. In the GEANT4 simulation, the evaluation of $f_{+1}$ is obtained from a Monte Carlo sampling of the charge state along the particle track according to the cross sections of $\sigma_{01}$, $\sigma_{10}$, $\sigma_{0-1}$ and $\sigma_{-10}$. As mentioned earlier, in Eq.~(\ref{eq:simple_formula}) only a part of the cross sections ($\sigma_{01}$ and $\sigma_{10}$) are used, neglecting $\sigma_{0-1}$ and $\sigma_{-10}$, since the latter values are orders of magnitude smaller.

\begin{figure}[!hbt]
	\centering
	\includegraphics[scale=0.4]{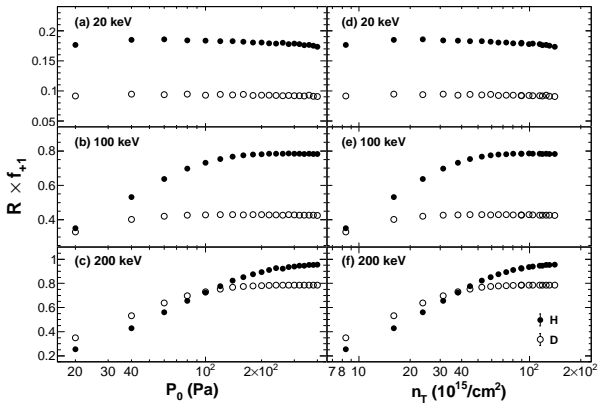}
	\caption{\footnotesize
		Global efficiency of $R \times f_{+1}$ as a function of the pressure $P_0$ for the incident energy $E$ = 20 keV (a), 100 keV (b) and 200 keV (c). (d), (e) and (f) are same as those of (a), (b) and (c) but as a function of $n_T$. Solid and open circles represent for H and D, respectively.
	}
	\label{fig:fig11_Ratio_times_Q1fraction}
\end{figure}

In order to determine the optimum condition, the global efficiency, a combination of transmission rate and fraction of charge state +1, $R \times f_{+1}$, is further studied. Fig.~\ref{fig:fig11_Ratio_times_Q1fraction} shows the global efficiency of $R \times f_{+1}$ as a function of $P_0$ for the incident energy $E$ = 20, 100 and 200 keV in (a), (b) and (c), and as a function of $n_T$ in (d), (e) and (f), respectively. The global efficiency of H and D decreases gradually as the pressure $P_0$ increases for $E$ = 20 keV. On the other hand, the global efficiency shows similar trend as that of $f_{+1}$ as the pressure $P_0$ increases for $E$ = 100 and 200 keV. For the pressure $P_0>$ 240 Pa, the global efficiency already becomes flat for all conditions of H and D at the incident energy $E\geq$ 100 keV. Considering the low temperature of the plasma in HL-2A/M, the number of high energy particles are in orders of magnitude less than that of the low energy particles. Therefore, $P_0 = $ 240 Pa is obtained as the optimum pressure for the maximum global efficiency in the incident energy range investigated. At this $P_0$, the pressure at the vacuum chamber is less than 0.1 Pa, which is inside the operating pressure range of the molecular pump. The simulation results would provide a useful guide for the actual applications.

\section{Summary} \label{sec.IV}

Neutral particle analyzer (NPA) is one of the crucial diagnostic devices on Tokamak facilities. The stripping unit is one of the main parts of NPA. A windowless gas tripping room with two differential pipes is adopted to keep a certain pressure for the parallel direction of electric and magnetic fields (E//B) NPA. The gas pressure distribution of the striping chamber is calculated, combining a computational fluid dynamics (CFD) software of Ansys Fluent and a Monte Carlo software of MolFlow+ for the low pressure collisionless region in the vacuum chamber. The pressure distribution along the beam direction is obtained for different input pressures. A certain high pressure is achieved inside the stripping room and the linear decreasing pressure is obtained inside the differential pipes and the entrance and exit holes. More than two orders magnitude smaller pressure is obtained in the vacuum chamber than that inside the stripping room. 

Base on the pressure distributions calculated by Ansys Fluent and MolFlow+, the stripping efficiency of the stripping chamber for H and D atoms at the incident energy ranging from 20 to 200 keV is studied using GEANT4. The energy loss of H and D after passing through the stripping chamber is small and can be neglected for all the incident energies and the input pressures investigated. The scattering loss of H and D atoms on H$_{2}$ gas is studied through the transmission rate ($R$) of the incident atoms. A slight increasing trend is observed for $R$ as the incident energy increases, but show an opposite trend as the input pressure ($P_0$) increases. The scattering loss is small (less than 3\%) for all the incident energies and the input pressures investigated. 

A charge state variable is introduced to track the charge state of particles in the GEANT4 simulation. Adopting the ORNL recommended charge exchange cross sections in a modified hadron reaction physics, the charge state of each particle is traced in the simulation. The behaviors of charge fractions along the beam direction (z-axis) in the H$_2$ gas are investigated for $E$ = 20, 100 and 200 keV H and D atoms. The stripping efficiency is obtained as the fraction of charge state +1 at the exit hole of the vacuum chamber (z = 120 mm). After reaching the maximum value, $f_{+1}$ decreases for $P_{0}$ = 20 Pa, and stay flat for $P_0$ = 100 Pa, but keep slowly increasing for $P_{0}$ = 400 Pa as the incident energy per nucleon increases. $f_{+1}$ quickly reach a maximum value and keep the maximum as $P_{0}$ increases at lower $E$. For larger $E$, the fractions increase faster at lower $P_{0}$ and reach the maximum at the input pressure around $P_{0}$ = 240 Pa. 

According to the combined global efficiency, $R \times f_{+1}$, $P_0$ = 240 Pa is found as the optimum pressure for the maximum global efficiency in the incident energy range investigated. The simulation results would provide a useful guide for the actual applications.

\end{document}